\documentclass[a4paper,11pt]{article}
\usepackage{pos}
\usepackage{hyperref}
\usepackage{subcaption}
\usepackage{dsfont}
\usepackage{slashed}

\newcommand{\m}{{\bf m}}
\newcommand{\h}{{\bf h}}

\usepackage{ifthen}
\usepackage{tikz}
\usepackage{xspace}

\newcommand{\dNa}{\delta \eta_{N_\alpha}}

\newcommand{\etaL}{\eta_L}
\newcommand{\za}{z_\alpha}
\newcommand{\mNa}{m_{N_\alpha}}

\newcommand{\zsph}{z_{\rm sph}}

\newcommand{\heff}{h_{\rm eff}}

\newcommand{\nn}{\nonumber}

\newcommand{\GeV}{{\rm GeV}\xspace}
\newcommand{\TeV}{{\rm TeV}\xspace}

\newcommand{\lrb}[1]{\left( #1 \right)}
\newcommand{\lrsb}[1]{\left[ #1 \right]}

\newcommand{\lrBiggcb}[1]{\Bigg\{ #1 \Bigg\}}
\newcounter{NumArgs}

\newcommand{\eqs}[1]{\setcounter{NumArgs}{0}\foreach\i in{#1}{\stepcounter{NumArgs}}%
	\ifthenelse{\equal{\theNumArgs}{1}}{(\ref{#1})}%
	{\ifthenelse{\equal{\theNumArgs}{2}}%
		{\foreach\i[count=\q]in{#1}{\ifthenelse{\equal{\q}{\theNumArgs}}{and (\ref{\i})}{(\ref{\i})~}}}%
		{\foreach\i[count=\q]in{#1}{\ifthenelse{\equal{\q}{\theNumArgs}}{and (\ref{\i})}{(\ref{\i}),~}}}}}

\newcommand{\Eqs}[1]{\setcounter{NumArgs}{0}\foreach\i in{#1}{\stepcounter{NumArgs}}%
	\ifthenelse{\equal{\theNumArgs}{1}}{Eq.~(\ref{#1})}%
	{\ifthenelse{\equal{\theNumArgs}{2}}%
		{Eqs.~\foreach\i[count=\q]in{#1}{\ifthenelse{\equal{\q}{\theNumArgs}}{and (\ref{\i})}{(\ref{\i})~}}}%
		{Eqs.~\foreach\i[count=\q]in{#1}{\ifthenelse{\equal{\q}{\theNumArgs}}{and (\ref{\i})}{(\ref{\i}),~}}}}}

\newcommand{\refs}[1]{\setcounter{NumArgs}{0}\foreach\i in{#1}{\stepcounter{NumArgs}}%
	\ifthenelse{\equal{\theNumArgs}{1}}{(\ref{#1})}%
	{\ifthenelse{\equal{\theNumArgs}{2}}%
		{\foreach\i[count=\q]in{#1}{\ifthenelse{\equal{\q}{\theNumArgs}}{and (\ref{\i})}{(\ref{\i})~}}}%
		{\foreach\i[count=\q]in{#1}{\ifthenelse{\equal{\q}{\theNumArgs}}{and (\ref{\i})}{(\ref{\i}),~}}}}}

\newcommand{\Figs}[1]{\setcounter{NumArgs}{0}\foreach\i in{#1}{\stepcounter{NumArgs}}%
	\ifthenelse{\equal{\theNumArgs}{1}}{Figure~\ref{#1}}%
	{\ifthenelse{\equal{\theNumArgs}{2}}%
		{Figures~\foreach\i[count=\q]in{#1}{\ifthenelse{\equal{\q}{\theNumArgs}}{and \ref{\i}}{\ref{\i}~}}}%
		{Figures~\foreach\i[count=\q]in{#1}{\ifthenelse{\equal{\q}{\theNumArgs}}{and \ref{\i}}{\ref{\i},~}}}}}

\newcommand{\Gen}[2]{\setcounter{NumArgs}{0}\foreach\i in{#2}{\stepcounter{NumArgs}}%
	\ifthenelse{\equal{\theNumArgs}{1}}{#1.~(\ref{#2})}%
	{\ifthenelse{\equal{\theNumArgs}{2}}%
		{#1.~\foreach\i[count=\q]in{#2}{\ifthenelse{\equal{\q}{\theNumArgs}}{and (\ref{\i})}{(\ref{\i})~}}}%
		{#1.~\foreach\i[count=\q]in{#2}{\ifthenelse{\equal{\q}{\theNumArgs}}{and (\ref{\i})}{(\ref{\i}),~}}}}}

\title{Tri-Resonant Leptogenesis}
\author{P. Candia da Silva}
\author{D. Karamitros}
\author{T. McKelvey}
\author*{A. Pilaftsis}

\affiliation{Department of Physics and Astronomy, University of Manchester,\newline Manchester, M13 9PL, United Kingdom}

\emailAdd{pablo.candiadasilva@manchester.ac.uk}
\emailAdd{dimitrios.karamitros@manchester.ac.uk}
\emailAdd{thomas.mckelvey@manchester.ac.uk}
\emailAdd{apostolos.pilaftsis@manchester.ac.uk}

\abstract{We present a class of leptogenesis models where the light neutrinos acquire their observed mass through a symmetry-motivated construction. We consider an extension of the Standard Model, which includes three singlet neutrinos which have mass splittings comparable to their decay widths. We show that this tri-resonant structure leads to an appreciable increase in the observed CP asymmetry over that found previously in typical bi-resonant models. To analyse such tri-resonant scenarios, we solve a set of coupled Boltzmann equations, crucially preserving the variations in the relativistic degrees of freedom. We highlight the fact that small variations at high temperatures can have major implications for the evolution of the baryon asymmetry when the singlet neutrino mass scale is below $100 \; \GeV$. We then illustrate how this variation can significantly affect the ability to find successful leptogenesis at these low masses. Finally, the parameter space for viable leptogenesis is delineated, and comparisons are made with current and future experiments.}

\FullConference{Corfu Summer Institute 2022 ``School and Workshops on Elementary Particle Physics and Gravity'', August 28 - September 8, 2022, Corfu, Greece.}

\begin{document}
\maketitle

%\newpage
%
\section{Introduction}\label{sec:Intro}
Observations done by the Wilkinson Microwave Anisotropy Probe (WMAP)
and the Planck observatory indicate that the extent of the Baryon
Asymmetry of the Universe (BAU) amounts
to~\cite{Planck:2018vyg,Fields:2019pfx} 
\begin{equation}
\eta_B^{\rm CMB} = 6.104\pm 0.058 \times 10^{-10}.
\end{equation}
Hence, explaining the observed BAU has been one of the central themes
of Particle Cosmology for decades. The existence of this non-zero BAU
is one of the greatest pieces of evidence for physics beyond the
Standard Model (SM). Moreover, the observation of neutrino oscillation phenomena \cite{Ahmad:2001an,Ahmad:2002jz,Fukuda:1998mi} indicates the existence of non-zero neutrino masses in contradiction with the SM prediction. A minimal resolution to both of these problems is to introduce additional neutrinos, which are singlets of the SM gauge group: $\textrm{SU}(3)_c \times \textrm{SU}(2)_L \times \textrm{U}(1)_Y$. The inclusion of a lepton number violating Majorana mass term permits these additional neutrinos to have large masses whilst suppressing the masses of the SM neutrinos. This mechanism is aptly referred to as the seesaw mechanism \cite{Minkowski:1977sc,GellMann:1980vs,Yanagida:1979as,Mohapatra:1979ia}. The violation of lepton number by two units satisfies one of the famous Sakharov conditions~\cite{Sakharov:1967dj} for the generation of appreciable particle asymmetries. Further to this, the expansion of the FRW Universe provides a cosmic arrow of time as well as satisfying the out-of-equilibrium condition, again provided by Sakharov. In combination with the CP violation present in the Yukawa sector, these properties allow for the generation of large lepton asymmetries, which may then be re-processed into a baryon asymmetry through equilibrium $(B+L)$-violating sphaleron transitions. This mechanism of generating appreciable BAU is widely known as \textit{leptogenesis}~\cite{FukYan:1986,Buchmuller:2004nz}.

In \cite{daSilva:2022mrx}, we consider a class of leptogenesis models which can provide naturally light SM neutrino masses as well as generate appreciable levels of BAU to match the observed CMB data. We assume these models to contain three singlet neutrinos, which have mass splittings comparable to their decay widths, permitting maximal mixing between the heavy eigenstates. This framework is commonly referred to as Resonant Leptogenesis (RL). To this end, we compute the CP asymmetries associated with this tri-resonant model and follow this up with a set of complete Boltzmann equations to calculate the generated BAU. These Boltzmann equations describe the evolution of the neutrino and lepton number densities prior to the sphaleron freeze-out when the temperature of the universe falls below $T_\textrm{sph}\approx 132 \; \GeV$~\cite{DOnofrio:2014rug}. A particular highlight of our study is the preservation of the variations in the relativistic degrees of freedom. In much of the literature, the degrees of freedom are taken to be constant due to the high-temperature scales. However, we show that the small variations, which pervade even above temperatures of $T = 100\;\GeV$, can have a significant impact on the generated BAU.

Finally, we present results for the allowed regions of parameter space which can achieve successful leptogenesis and compare these results with current and future experiments. In particular, we make comparisons with coherent flavour-changing processes within nuclei from experiments such as MEG~\cite{MEG:2016leq,MEGII:2018kmf} and PRISM~\cite{BARLOW201144}, as well as collider experiments such as the LHC and FCC.

\section{Flavour Symmetric Model}\label{sec:Model}
We utilise a minimal extension of the SM, with the inclusion of three right-handed neutrinos, which are singlets of the SM gauge group: $\textrm{SU}(3)_c \times \textrm{SU}(2)_L \times \textrm{U}(1)_Y$, and have lepton number $\textrm{L} = 1$. Given this additional particle content and quantum number assignment, the SM is extended through the additional Lagrangian terms
\begin{align}
    \mathcal{L}_{\nu_R} =i\overline{\nu}_R\slashed{\partial}\nu_R -\left( \overline{L}\,\h^\nu\tilde{\Phi}\,\nu_R + \frac{1}{2}\overline{\nu}^C_R\,\m_M\nu_R + {\rm H.c.\, }\right)\label{eq:seesaw_lagrangian}.
\end{align}
Here, $L_i = \left(\nu_{L, i}, e_{L, i} \right)^{\sf T}$, with $i=1,2,3$, are left-handed lepton doublets; $\nu_{R, \alpha}$, with $\alpha=1,2,3$, are right-handed neutrino singlet fields; and $\tilde{\Phi}$ is the isospin conjugate Higgs doublet. The matrices $\mathbf{h}_{i\alpha}^\nu$ and $(\mathbf{m}_M)_{\alpha\beta}$ are the neutrino Yukawa couplings and Majorana mass matrix, respectively. It is worth pointing out that the inclusion of the Majorana mass matrix explicitly breaks lepton number conservation by two units, $\Delta L = 2$, satisfying one of the three Sakharov conditions for the generations of appreciable lepton asymmetry.

Without loss of generality, we may select a basis for the singlet neutrino sector such that the Majorana mass term is diagonalised, \textit{i.e} $\m_M = \textrm{diag}(m_{N_1}, m_{N_2}, m_{N_3})$. In this basis, the Lagrangian in the unbroken phase takes the form
\begin{align}
\mathcal{L}_{\nu_R} =i\overline{N}\slashed{\partial}N-\left( \overline{L}\,\h^\nu\tilde{\Phi}\,P_R N + {\rm H.c.}\right) - \frac{1}{2}\overline{N}\,\m_MN.\label{eq:seesaw_lagrangian_mass}
\end{align}
In this expression, $N_\alpha = \nu_{R, \alpha} + \nu_{R, \alpha}^C$, and $P_{R/L} = \frac{1}{2}\left(\mathds{1}_4 \pm \gamma^5 \right)$ are right/left-chiral projection operators.

In the broken phase, the addition of a Dirac mass term from the Yukawa sector results in the mixing between left- and right-chiral neutrinos, with the mass basis in the broken phase a particular combination of left- and right-chiral neutrinos
\begin{align}
P_R \begin{pmatrix}
\nu \\
N
\end{pmatrix}=
\begin{pmatrix}
U_{\nu\nu_L^C} & U_{\nu \nu_R}\\
U_{N\nu_L^C} & U_{N \nu_R}
\end{pmatrix}
\begin{pmatrix}
\nu_L^C\\
\nu_R
\end{pmatrix}\;.
\end{align}
In the above, we have defined $\nu_i$ as light neutrino mass eigenstates and $N_i$ as heavy neutrino mass eigenstates. Furthermore, the unitary matrix, $U$, diagonalises the full neutrino mass matrix. To leading order in the quantity $\xi_{i\alpha} = (\mathbf{m}_D \mathbf{m}_M^{-1})_{i\alpha}$~\cite{Pilaftsis:1991ug}, the light neutrino mass matrix may be written as
\begin{align}
\m^\nu = -\m_D \m^{-1}_M \m^{\sf T}_D \; ,\label{eq:tree_level_mass}
\end{align}
with $\m_D = \mathbf{h}^\nu v /\sqrt{2}$ the Dirac mass matrix, with Higgs VEV, $v \simeq 246\; \GeV$~\cite{GellMann:1980vs}. By virtue of this relation, it is clear that to satisfy observed neutrino data, the Majorana mass matrix would have to be GUT scale if the Dirac matrix is of electroweak scale ($|\!|\m_D|\!| \sim v$), and of general structure. As a result, the impact of singlet neutrinos on experimental signatures would be minimal as the charged current interactions are suppressed through the mixing parameter $B_{i\alpha} = \xi_{i\alpha}$~\cite{Pilaftsis:1991ug}
\begin{align}
\mathcal{L}^W_{\rm int} = -\frac{g_w}{\sqrt{2}}W^-_\mu
  \overline{e}_{iL} B_{i\alpha}\gamma^\mu P_L N_\alpha + {\rm H.c.}\,.
\end{align}
Consequently, there is a motivation to identify models which allow for low-scale heavy neutrino masses whilst remaining in alignment with the observed neutrino data.

One approach which may be taken to address this problem is to assume the existence of a symmetry on the flavour structure of the Yukawa couplings, $\mathbf{h}_0^\nu$, which would render the light neutrino eigenstates massless
\begin{equation}
    -\m_D \m^{-1}_M \m^{\sf T}_D = -\frac{v^2}{2}\mathbf{h}_0^\nu \m^{-1}_M (\mathbf{h}_0^\nu)^{\sf T} = \mathbf{0}_3.\label{eq:zeromassconstraint}
\end{equation}
From this, small neutrino masses may be generated through perturbations about the symmetric Yukawa couplings
\begin{align}
    (\h^\nu_0 + \delta\h^\nu)\,\m_M^{-1}\,(\h^\nu_0 + \delta\h^\nu)^{\sf
  T}\, =\, \frac{2}{v^2}\,\m^\nu\;. 
    \label{eq:numassconstraint}
\end{align}

In the case of a near degenerate heavy neutrino mass spectrum, the condition on the symmetric Yukawa couplings given in (\ref{eq:zeromassconstraint}) may be approximately satisfied by
\begin{equation}
    \mathbf{h}_0^\nu (\mathbf{h}_0^\nu)^{\sf T} =\mathbf{0}_3.\label{eq:NilPotent}
\end{equation}
This motivates a nil-potent structure of the Yukawa matrix. In particular, we have identified the structure
\begin{align}
    \h^\nu_0=\begin{pmatrix}
        a & a\,\omega & a\,\omega^2\\
        b & b\,\omega & b\,\omega^2\\
        c & c\,\omega & c\,\omega^2
    \end{pmatrix}\;,
    \label{eq:z6yukawa}
\end{align}
with $a, \, b, \,c \in \mathbb{C}$, and $\omega = \exp\left( \frac{2\pi i}{6} \right)$ the generator of the $\mathbb{Z}_6$ group. This structure is not unique in satisfying the constraint given in (\ref{eq:NilPotent}). Other similar structures, such as $\mathbb{Z}_3$, with generators $\omega^\prime = \exp\left( \frac{2\pi i}{3} \right)$ would also produce a vanishing light neutrino mass spectrum at leading order. Most interestingly, this symmetry-motivated structure offers large CP-violating phases which contribute significantly to the generation of appreciable BAU.
Instead, this possibility is not easily achievable in bi-resonant models, where the CP-odd phases are strongly correlated to the light-neutrino masses.

As a further insight, since this symmetry exists within the flavour structure, any additional contributions to the light neutrino mass matrix with an identical flavour structure will vanish. In particular, the first-order loop correction to the light neutrino mass matrix~\cite{Pilaftsis:1991ug} may be incorporated into the zero mass condition of the symmetric Yukawa matrix
\begin{align}
    \frac{v^2}{2}\h^\nu_0\left[\m^{-1}_M - \frac{\alpha_w}{16\pi M^2_W}\m^{\dagger}_Mf(\m_M\m^\dagger_M)\right]\h^{\nu\sf
  T}_0=\,\mathbf{0}_3\;,
    \label{eq:one_loop_zero_mass}
\end{align}
where
\begin{align}
    f(\m_M\m^\dagger_M)=\frac{M^2_H}{\m_M\m^\dagger_M -
M^2_H\mathds{1}_3}\ln\left(\frac{\m_M\m^\dagger_M}{M^2_H}\right) +
  \frac{3M^2_Z}{\m_M\m^\dagger_M -
M^2_Z\mathds{1}_3}\ln\left(\frac{\m_M\m^\dagger_M}{M^2_Z}\right)\;. 
    \label{eq:loop_factor_f_def}
\end{align}
In the above, $\alpha_w\equiv g_w^2/(4\pi)^2$ is the electroweak gauge-coupling parameter, and $M_W$, $M_Z$, and $M_H$ are the masses of the $W$, $Z$, and Higgs bosons, respectively.

\section{Leptonic Asymmetries}\label{sec:CPA}
In models of thermal leptogenesis, CP-violating effects enter through the difference in the decay rates of heavy neutrinos into leptons and Higgs bosons $(N \rightarrow L \Phi)$, and the conjugate process $(N \rightarrow L^c \Phi^\dagger)$~\cite{FukYan:1986,Buchmuller:2004nz}. This difference appears at the loop level, with the wavefunction contribution particularly dominant in models of RL, where mass splittings are of a similar size to the decay widths of the heavy neutrinos (for a review, see~\cite{Pilaftsis:1998pd}). To aid the discussion of analytic results regarding the leptonic CP asymmetries, we introduce the coefficients~\cite{Pilaftsis:1997jf,Pilaftsis:2003gt,Pilaftsis:2005rv}
\begin{align}
    A_{\alpha \beta} &= \sum_{l=1}^3 \frac{\h^\nu_{l \alpha}\h_{l
                       \beta}^{\nu*}}{16\pi} =
                       \frac{(\h^{\nu\dagger}\h^\nu)^*_{\alpha\beta}}{16\pi},\\ 
    V_{l \alpha} &= \sum_{k=1}^3 \sum_{\gamma \neq \alpha}
                   \frac{\h^{\nu*}_{k\alpha}\h^\nu_{k
                   \gamma}\h^\nu_{l\gamma}}{16\pi} f\left(
                   \frac{m^2_{N_\gamma}}{m^2_{N_\alpha}} \right),
    \label{eq:V_def}
\end{align}
which correspond to absorptive transition rates for the wavefunction and vertex, respectively. In~(\ref{eq:V_def}), ${f(x) = \sqrt{x}\left[1-(1+x)\ln \left( \frac{1+x}{x} \right) \right]}$ is the Fukugita-Yanagida 1-loop function~\cite{FukYan:1986,Buchmuller:2004nz}.

Completing a full re-summation of the loop corrections, including all three Majorana neutrinos, generates an effective $NL\tilde{\Phi}$ coupling~\cite{Pilaftsis:2003gt, Pilaftsis:2005rv, Deppisch:2010fr}
\begin{align}
\label{eq:Eff_Yuk}
    (\bar{\mathbf{h}}^\nu_+)_{l\alpha} =&\; \h^\nu_{l\alpha} +
                                          iV_{l\alpha} - i
                                          \sum_{\beta,\gamma = 1}^3
                                          |\varepsilon_{\alpha\beta\gamma}|\,\h^\nu_{l\beta}\nonumber\\&\times
  \frac{m_{N_\alpha}\left(M_{\alpha\alpha\beta}+M_{\beta\beta\alpha}\right)-i
  R_{\alpha\gamma}
  \left[M_{\alpha\gamma\beta}\left(M_{\alpha\alpha\gamma}+M_{\gamma\gamma\alpha}\right)
  +
  M_{\beta\beta\gamma}\left(M_{\alpha\gamma\alpha}+M_{\gamma\alpha\gamma}\right)\right]}{m_{N_\alpha}^2-m_{N_\beta}^2
  + 2i m^2_{N_\alpha} A_{\beta\beta} + 2i\,\Im m
  R_{\alpha\gamma}\left(m_{N_\alpha}^2 |A_{\beta\gamma}|^2 +
  m_{N_\beta} m_{N_\gamma} \Re e A_{\beta\gamma}^2 \right)}\;, 
\end{align}
where $\epsilon_{\alpha\beta\gamma}$ is the anti-symmetric Levi-Civita
symbol, $M_{\alpha\beta\gamma}\equiv m_{N_\alpha}A_{\beta\gamma}$ and 
\begin{equation}
    R_{\alpha\beta} \equiv \frac{m_{N_\alpha}^2}{m_{N_\alpha}^2-m_{N_\beta}^2+2i m_{N_\alpha}^2 A_{\beta\beta}}\;.
\end{equation}
The conjugate $NL^c\tilde{\Phi}^\dagger$ couplings, denoted by $(\bar{\mathbf{h}}^\nu_-)_{l\alpha}$, are found through the replacement of $\mathbf{h}^\nu_{l\alpha}$ by $(\mathbf{h}^\nu)^*_{l\alpha}$ in (\ref{eq:Eff_Yuk}). These effective couplings capture both \textit{bi-resonant} and \textit{tri-resonant} effects, corresponding to maximal CP asymmetries through the mixing of two and three singlet neutrinos, respectively. In particular, one may recover the bi-resonant expressions by simply taking $R_{\alpha\gamma}$ to~zero.

Utilising these re-summed effective couplings, we may calculate the partial decay widths of the heavy neutrinos as
\begin{equation}
    \Gamma (N_\alpha \rightarrow L_l \Phi) =
    \frac{m_{N_\alpha}}{8\pi}\left| (\bar{\h}^\nu_+)_{l \alpha}
    \right|^2,\qquad 
    \Gamma (N_\alpha \rightarrow L^C_l \Phi^\dagger) =
    \frac{m_{N_\alpha}}{8\pi}\left| (\bar{\mathbf{h}}^\nu_-)_{l
        \alpha} \right|^2\;. 
    \label{eq:Gammas_def}
\end{equation}

\begin{figure}[t!]
\hspace*{-0.6cm}
%\centering
    \includegraphics[width=\linewidth]{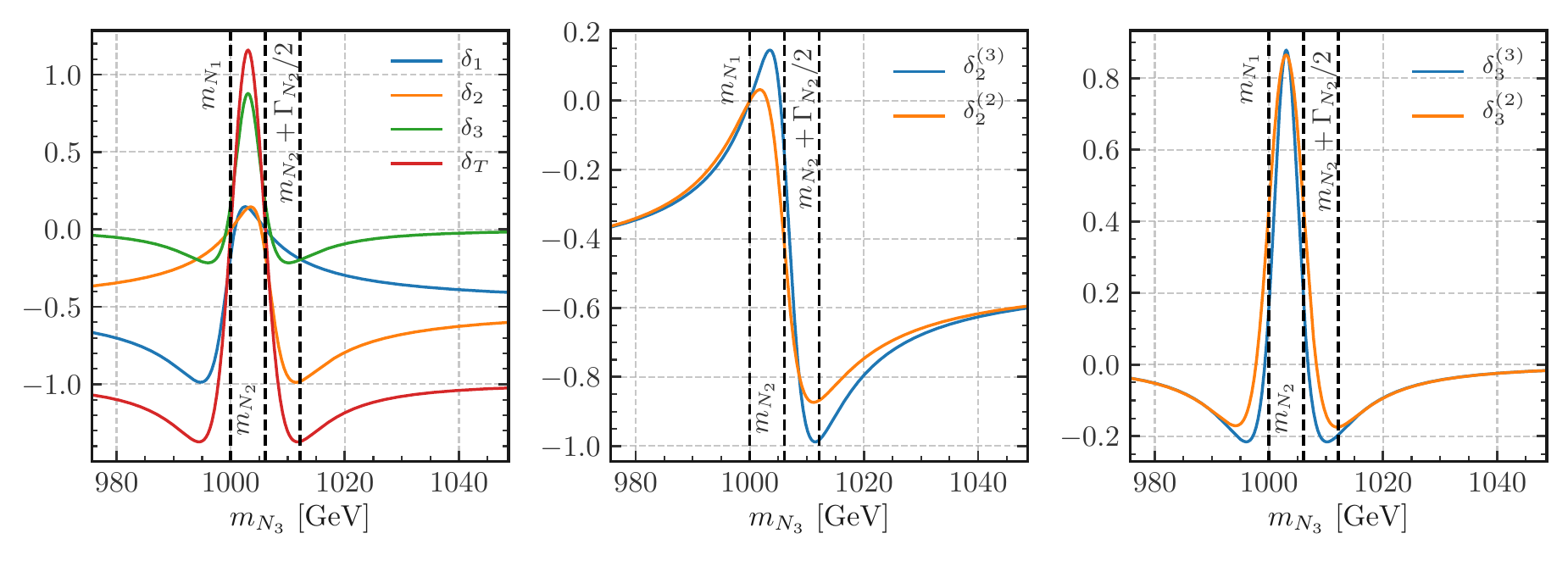}
    \caption{\textit{Left panel:} CP asymmetries generated by the decays of
      $N_1$, $N_2$ and $N_3$, together with the total CP asymmetry
      $\delta_T = \sum_\alpha \delta_\alpha$, as a function of the
      mass of $N_3$. \textit{Centre panel:} Comparison of the CP asymmetry in the decay
      of $N_2$ vs. $m_{N_3}$ as calculated from two neutrino mixing ($\delta^{(2)}_2$) and three-neutrino mixing
      ($\delta^{(3)}_2$). \textit{Right panel:} CP asymmetry in the
      decay of $N_3$ vs. $m_{N_3}$ calculated from two-neutrino mixing ($\delta^{(2)}_3$) and
      three-neutrino mixing ($\delta^{(3)}_3$). We indicate the
      values of $m_{N_1}$, $m_{N_2}$ and the tri-resonant value of
      $m_{N_3}$ with grey dashed lines.} 
    \label{fig:cp_asymmetry}
\end{figure}

From this, we identify the size of the CP asymmetries within the model using the dimensionless quantity
\begin{equation}
    \delta_{\alpha l} \equiv \frac{\Gamma (N_\alpha \rightarrow L_l
      \Phi) - \Gamma (N_\alpha \rightarrow L^C_l \Phi^\dagger) }{
      \sum_{k=e,\mu,\tau} \Gamma (N_\alpha \rightarrow L_k \Phi)
      +\Gamma (N_\alpha \rightarrow L^C_k \Phi^\dagger)} =
    \frac{\left| (\bar{\mathbf{h}}^\nu_+)_{l \alpha} \right|^2 -
      \left| (\bar{\mathbf{h}}^\nu_-)_{l \alpha}
      \right|^2}{(\bar{\mathbf{h}}^{\nu\dagger}_+\bar{\mathbf{h}}^\nu_+)_{\alpha\alpha}
      +
      (\bar{\mathbf{h}}^{\nu\dagger}_-\bar{\mathbf{h}}^\nu_-)_{\alpha\alpha}}. 
    \label{eq:CP-deltas_def}
\end{equation}
Furthermore, we may define the total CP asymmetry associated with each neutrino species by summing over the lepton families
\begin{equation}
    \delta_\alpha = \sum_{l} \delta_{\alpha l} = 
    \frac{(\bar{\mathbf{h}}^{\nu\dagger}_+\bar{\mathbf{h}}^\nu_+)_{\alpha\alpha}
      -
      (\bar{\mathbf{h}}^{\nu\dagger}_-\bar{\mathbf{h}}^\nu_-)_{\alpha\alpha}}{(\bar{\mathbf{h}}^{\nu\dagger}_+\bar{\mathbf{h}}^\nu_+)_{\alpha\alpha}
      +
      (\bar{\mathbf{h}}^{\nu\dagger}_-\bar{\mathbf{h}}^\nu_-)_{\alpha\alpha}}.
\end{equation}
At this point, it is important to mention that the existence of non-zero CP asymmetries is only possible in the event that the CP-odd invariant
\begin{align}
    \Delta_{\rm CP} &= \Im m \left\{ \textrm{Tr} \left[
                  (\mathbf{h}^\nu)^\dagger \mathbf{h}^\nu
                  \mathbf{m}_M^\dagger \mathbf{m}_M
                  \mathbf{m}_M^\dagger (\mathbf{h}^\nu)^{\sf T}
                  (\mathbf{h}^\nu)^* \mathbf{m}_M\right] \right\}\\ 
    &= \sum_{\alpha<\beta} m_{N_\alpha} m_{N_\beta}
      \left(m_{N_\alpha}^2 - m_{N_\beta}^2 \right)\, \Im m \Big[
      \left(\mathbf{h}^{\nu\dagger}\mathbf{h}^\nu
      \right)_{\beta\alpha}^2 \Big]
\end{align}
does not vanish~\cite{Pilaftsis:1997dr, Pilaftsis:2003gt,
Branco:1986gr, Yu:2020gre} When all neutrinos are exactly degenerate, this quantity is trivially zero, and hence CP asymmetries are not possible. However, if mass splittings are permitted, we see that in the $\mathbb{Z}_6$ model presented earlier, this CP-odd quantity is proportional to the $\mathbb{Z}_6$ element $\omega^2$
\begin{align}
    \Delta_{\rm CP} &\approx \left(|a|^2 + |b|^2 + |c|^2 \right)^2
                  \sum_{\alpha<\beta} m_{N_\alpha} m_{N_\beta}
                  \left(m_{N_\alpha}^2 - m_{N_\beta}^2 \right)\, \Im m
                  \Big( \omega^{2(\alpha-\beta)} \Big)\; . 
\end{align}
Accordingly, the $\mathbb{Z}_6$ structure we have proposed offers both naturally light SM neutrino masses and produce significant levels of CP asymmetry due to the large CP-violating phases present.

In the literature, there are several examples of the bi-resonant approximation being used in RL scenarios to enhance the contribution to the CP asymmetry through the mixing of two singlet neutrinos whilst permitting the third neutrino to decouple, either through suppressed couplings or a higher mass scale. However, in a model where all three neutrinos satisfy the resonance condition
\begin{equation}
    |m_{N_\alpha} - m_{N_\beta}| \sim \frac{\Gamma_{N_{\alpha,\beta}}}{2},
\end{equation}
the contributions to CP asymmetries may be enhanced through constructive interference between all three neutrinos~\cite{Pilaftsis:1997dr}. Figure~\ref{fig:cp_asymmetry} shows the variation in the generated CP asymmetry through the decay of singlet neutrinos, as well as the total CP asymmetry $\delta_T = \sum_\alpha \delta_\alpha$. In this figure, $m_{N_1}$ and $m_{N_2}$ are fixed to satisfy the resonance condition, and $m_{N_3}$ is permitted to vary. As is expected, the total CP asymmetry is seen to vanish in the case that $m_{N_3}$ is equal to either $m_{N_1}$ or $m_{N_2}$; however, is maximised when $m_{N_3} = m_{N_2} + \frac{1}{2}\Gamma_{N_2}$. This maximum of the CP asymmetry is $35\%$ larger than what can be produced in models with only two neutrino mixing. Furthermore, at this maximum, it can be seen that $\delta_1 \simeq \delta_3$, while $\delta_2$ is significantly enhanced. Consequently, $\delta_2$ is the dominant contributor to $\delta_T$.

The latter two panels in Figure~\ref{fig:cp_asymmetry} highlight the difference between two neutrino mixing, $\delta^{(2)}_\alpha$, and three neutrino mixing $\delta^{(3)}_\alpha$. It is clear from the second panel that the proper inclusion of three neutrino mixing is important in the resonant region, as a sizeable difference becomes apparent in the CP asymmetry of $N_2$. A similar effect is present in the CP asymmetry of $N_3$, shown in the final panel, although to a lesser extent.

In general, Figure~\ref{fig:cp_asymmetry} highlights the importance of full and proper accounting for the mixing of three neutrinos when these neutrinos are in consecutive resonance. As a consequence, this tri-resonant structure saturates the available CP asymmetry and maximises the generated BAU at a given mass scale with specified couplings. This is in contrast to the bi-resonant models commonly studied in the literature, which neglect contributions to the CP asymmetry from the mixing of a third neutrino species. 

\section{Boltzmann Equations}\label{sec:BEs}
The generation of appreciable BAU requires not only significant CP asymmetries but also a departure from equilibrium and baryon number violation. Here, we will introduce a complete set of Boltzmann equations which describe the out-of-equilibrium dynamics in the early universe, which allows for a dynamical generation of appreciable lepton asymmetry. This lepton asymmetry may be reprocessed into a baryon asymmetry through $(B+L)$-violating sphaleron transitions~\cite{KUZMIN198536}.

At temperature scales pertinent to leptogenesis, it is assumed that the Universe is in the radiation-domination era, with energy and entropy densities
\begin{equation}
    \rho(T) = \frac{\pi^2}{30} g_\textrm{eff}(T)\, T^4, \qquad s(T) = \frac{2\pi^2}{45} h_\textrm{eff}(T)\, T^3,
\end{equation}
respectively. Here, $T$ is the temperature of the Universe, with $g_\textrm{eff}$ and $h_\textrm{eff}$ relativistic degrees of freedom of the SM plasma. We include the variations in the relativistic degrees of freedom since these are not constant, even when the temperature is well above $100\; \GeV$. These variations are small in magnitude but may have drastic implications for the generation of appreciable BAU with low-scale neutrino masses. For our numerical simulations, we utilise the data set labelled `EOS C' provided in \cite{Hindmarsh:2005ix}.

The evolution of the neutrino and lepton asymmetry number densities are described by their respective Boltzmann equations, written as a function of the dimensionless parameter $z_\alpha = m_{N_\alpha}/T$. To align with previously used conventions, we define $z=z_1$ to be the dynamical evolution parameter. In addition, we normalise the number density of a species, $i$, to the photon density,
\begin{equation}
    n_{\gamma}(\za) = \frac{2\zeta(3)T^3}{\pi^2} = \frac{2\zeta(3)}{\pi^2} \lrb{\dfrac{\mNa}{\za}}^3.
\end{equation}
This normalisation simplifies the Boltzmann equations and relates the number density to an observable quantity,
\begin{equation}
    \eta_i(\za) = \frac{n_i(\za)}{n_\gamma(\za)}.
\end{equation}
In the case of the neutrino Boltzmann equations, it is convenient to express the evolution in terms of a departure-from-equilibrium quantity
\begin{equation}
    \delta\eta_\alpha(\za) = \frac{\eta_\alpha(\za)}{\eta_\alpha^{\textrm{eq}}(\za)} - 1.
\end{equation}
In this definition, we have used the equilibrium value of $\eta_\alpha$, which may be explicitly calculated to be
\begin{equation}
    \eta_\alpha^{\textrm{eq}} \approx \frac{\za^2}{2\zeta(3)} K_2(\za),
\end{equation}
with $\zeta(3)$ Ap\'ery's constant, and $K_n(\za)$ a modified Bessel function of the second kind.

With these considerations, we may write a set of coupled Boltzmann equations, including decay terms, $\Delta L = 1$ and $\Delta L = 2$ scattering processes, as well as the running of the degrees of freedom,
\begin{align}
    \frac{d\dNa}{d \ln \za} =&-\dfrac{ \delta_h(\za)}{H(\za) \ \eta_{N_\alpha}^{\rm eq}(\za) }\lrsb{\dNa  \lrb{\Gamma^{D(\alpha)} + \Gamma^{S(\alpha)}_Y + \Gamma^{S(\alpha)}_G} +\frac{2}{9}\, \etaL\, \delta_\alpha \lrb{\tilde\Gamma^{D(\alpha)} + \hat{\Gamma}^{S(\alpha)}_Y + \hat{\Gamma}^{S(\alpha)}_G} } \nonumber\\
    & +\lrb{\dNa+1} \, \left[\za \frac{K_1(\za)}{K_2(\za)} - 3(\delta_h(\za) -1) \right]\;,
    \label{eq:BEN}\\
    \frac{d\eta_L}{d \ln z} =& -  \frac{\delta_h(z)}{H(z)} \lrBiggcb {\sum_{\alpha=1}^3 \dNa \delta_\alpha \lrb{ \Gamma^{D(\alpha)} + \Gamma^{S(\alpha)}_Y + \Gamma^{S(\alpha)}_G} \nonumber \\
    & +\frac{2}{9}\eta_L \left[\sum_{\alpha=1}^3\left(\tilde\Gamma^{D(\alpha)} + \tilde{\Gamma}^{S(\alpha)}_Y + \tilde\Gamma^{S(\alpha)}_G + 
    \Gamma^{W(\alpha)}_Y + \Gamma^{W(\alpha)}_G\right) + \Gamma^{\Delta L = 2} \right] \nn\\ 
    & + \dfrac{2}{27} \etaL \, \sum_{\alpha=1}^3 \delta_\alpha^2 \, \lrb{ \Gamma^{W(\alpha)}_Y + \Gamma^{W(\alpha)}_G } }
    - 3\etaL(\delta_h(z) -1) \label{eq:BEL}\;.
\end{align}
In these equations, we have utilised the well-known expression for the temperature-dependent Hubble parameter
\begin{equation}
    H(\za) = \sqrt{\frac{4\pi^3 g_{\textrm{eff}}(\za)}{45}}\frac{m_{N_\alpha}^2}{M_{\textrm{Pl}}} \frac{1}{\za^2},
\end{equation}
with $M_{\textrm{Pl}} \approx 1.221 \times 10^{19} \; \GeV$ the Planck mass. The relevant collision terms, denoted by $\Gamma^X_Y$, are readily available in the literature \cite{Pilaftsis:2005rv}.

As presented here, the Boltzmann equations are complete up to $\Delta L = 2$ scattering processes. Moreover, the terms included take into account the subtraction of real intermediate states (RIS). Such terms may contribute negatively to the Boltzmann equations due to the lack of an on-shell contribution to the scattering amplitude.

As briefly alluded to earlier, the lepton asymmetry generated is partially re-processed into a baryon asymmetry through $(B+L)$-violating sphaleron transitions. As may be found in the literature \cite{PhysRevD.42.3344}, the generated BAU from a lepton asymmetry is given by
\begin{equation}
    \eta_B = -\frac{28}{51} \eta_L.
\end{equation}
However, sphaleron transitions become suppressed once the temperature of the Universe falls below the critical temperature $T_\textrm{sph} \approx 132 \; \GeV$. Consequently, no leptons are re-processed once the Universe cools to a temperature below $T_\textrm{sph}$. 

Moreover, observations of the BAU produce the value at the recombination epoch; however, due to the expansion of the Universe, this results in a dilution of the overall baryon asymmetry present at $T_\textrm{sph}$. To compare these two values meaningfully, we assume that as the Universe cools from $T_\textrm{sph}$ to $T_\textrm{rec}$, there are no entropy-releasing processes. Consequently, the entropy of the Universe is constant, and we may calculate the ratio
\begin{equation}
    \frac{\eta_B(T_\textrm{rec})}{\eta_B(T_\textrm{sph})} =\frac{n_\gamma(T_\textrm{sph}) s(T_\textrm{rec})}{n_\gamma(T_\textrm{rec}) s(T_\textrm{sph})} = \frac{h_{\textrm{eff}}(T_\textrm{rec})}{h_{\textrm{eff}}(T_\textrm{sph})}.
\end{equation}
For this ratio, we take the approximate value $1/27$~\cite{Pilaftsis:2003gt, Buchmuller:2004nz}, and as a result, the observed baryon asymmetry is related to the generated lepton asymmetry by
\begin{equation}
    \eta_B^\textrm{obs} = - \frac{1}{27}\frac{28}{51} \eta_L(T_\textrm{sph}).
\end{equation}

\section{Variations in the Relativistic Degrees of Freedom of the SM Plasma}
 \begin{figure}[t!]
    \centering
    \begin{subfigure}[]{0.48\textwidth}
    \includegraphics[width=1.\textwidth]{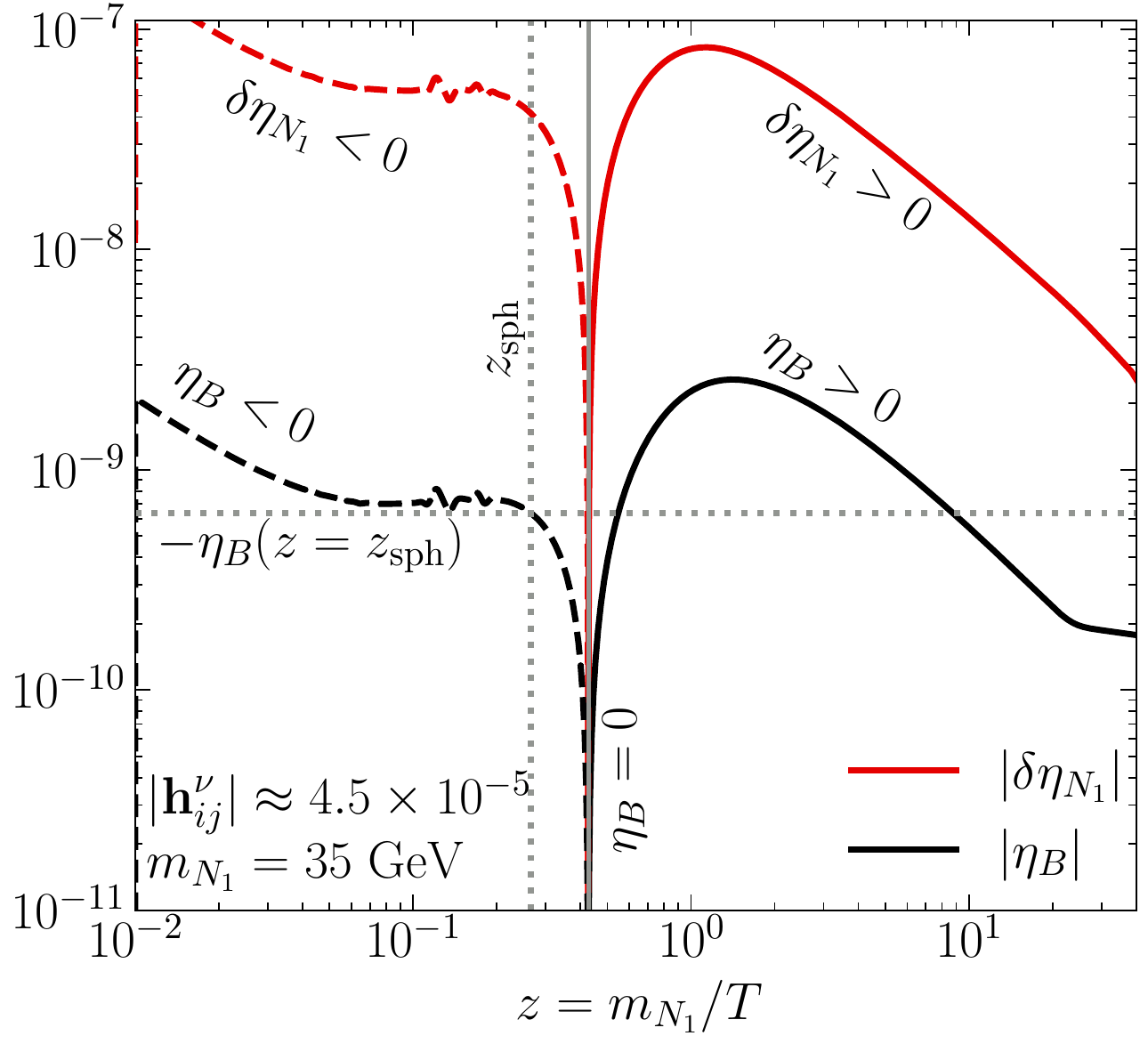}
    \caption{}
    \label{fig:etaB_negative}
    \end{subfigure}
    \begin{subfigure}[]{0.48\textwidth}
    \includegraphics[width=1.\textwidth]{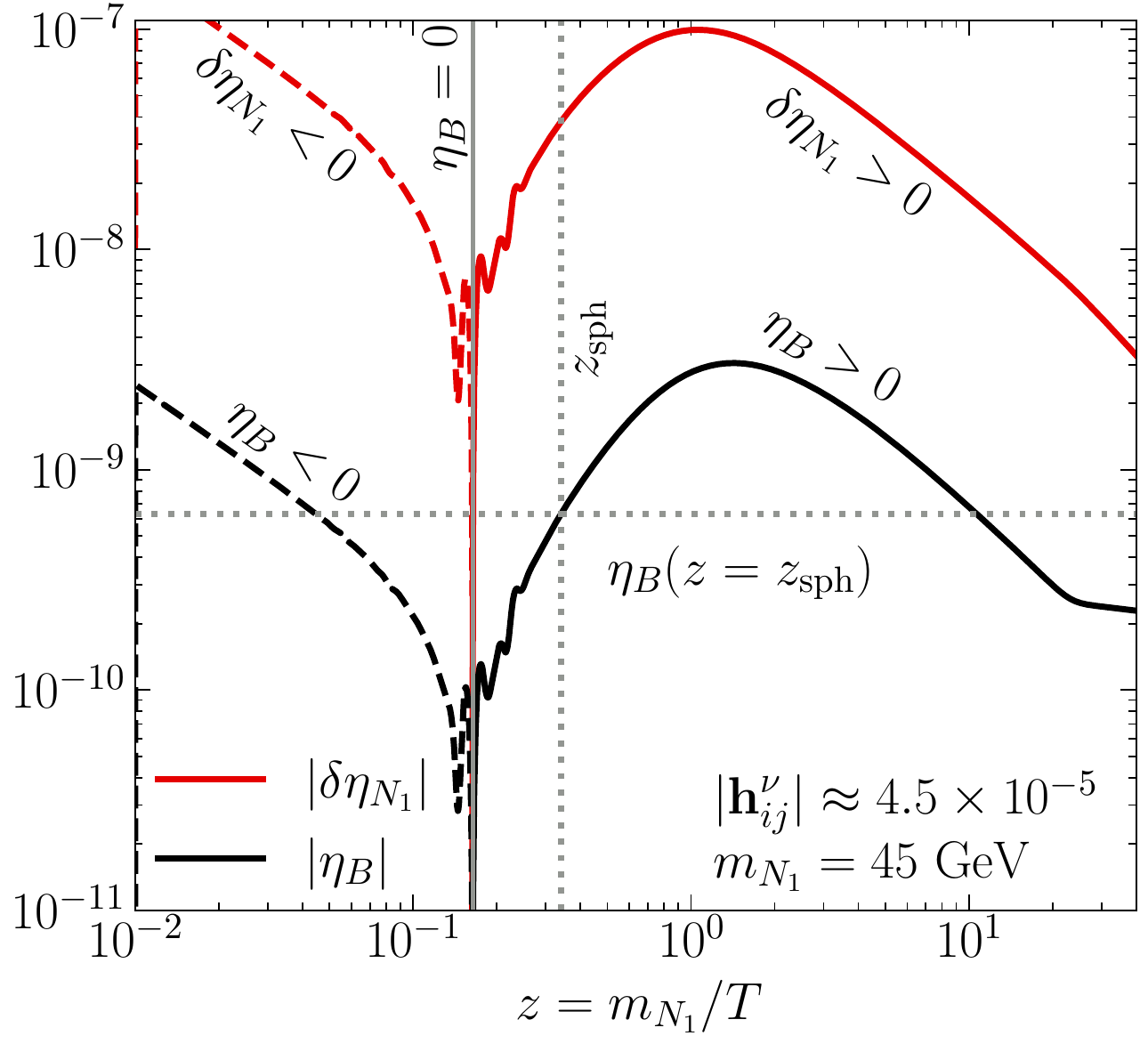}
    \caption{}
    \label{fig:etaB_positive}
    \end{subfigure}
    \caption{Evolution of $\delta_{N_1}$ (red) and $\etaL$ (black) for
      $m_{N_1} = 35~\GeV$ (a) and $m_{N_1} = 45~\GeV$ (b), with
      $|\h^{\nu}_{ij}| \approx 4.5 \times 10^{-5}$. The black and red
      solid (dashed) lines indicate where $\eta_B$ and $\delta_{N_1}$ are
      positive (negative). Grey dotted lines indicate the sphaleron freeze-out value, $z_\textrm{sph}$, and the observed Baryon asymmetry at $z_\textrm{sph}$.} 
    \label{fig:etaB_sign}
\end{figure}

In this section, we consider the effect of variations in the relativistic degrees of freedom of the SM plasma. In the Boltzmann equations, we introduce this effect through the inclusion of the factor
\begin{equation}
    \delta_h(z) = 1 - \frac{1}{3}\frac{d\ln h_{\textrm{eff}}}{d \ln z},\label{eq:DoFs}
\end{equation}
which is greater than 1 and has the limiting value $1$ for constant degrees of freedom.

While the size of this quantity does not differ significantly from unity, the final term of (\ref{eq:BEN}) may be dominant for low values of $z$. Consequently, early in the evolution of $\delta\eta_{N_\alpha}$, negative values may be observed. Accordingly, the dependence of $\eta_L$ on $\delta\eta_{N_\alpha}$ will result in negative values for the baryon asymmetry, $\eta_B$. In the case that the singlet neutrino mass scale is sufficiently low, the sphaleron freeze-out at $T_\textrm{sph}$ may preserve this feature.

In Figure~\ref{fig:etaB_sign}, we can see the impact of the variations in the relativistic degrees of freedom. In these figures, the dashed lines represent regions where the relevant quantity takes on negative values, typically $z \lesssim 10^{-1}$, and solid lines represent positive values, typically $z \gtrsim 10^{-1}$. Whilst these two figures are qualitatively similar, the importance of the mass scale becomes clear when we consider the values at the sphaleron freeze-out temperature. In Figure~\ref{fig:etaB_negative}, we see that for singlet neutrinos of mass $m_{N_1} = 35 \; \GeV$, the sphaleron freeze-out occurs prior to the evolution `bouncing back' to positive values, resulting in an overall negative sign for the generated BAU. Conversely, in Figure~\ref{fig:etaB_positive}, we consider singlet neutrinos of mass $m_{N_1} = 45 \; \GeV$, the baryon asymmetry has had enough time to return to positive values before the sphaleron freeze-out, and we find the expected positive values for the generated BAU.

\begin{figure}[t!]
\centering
	\begin{subfigure}[]{0.48\textwidth}
		\includegraphics[width=1.\textwidth]{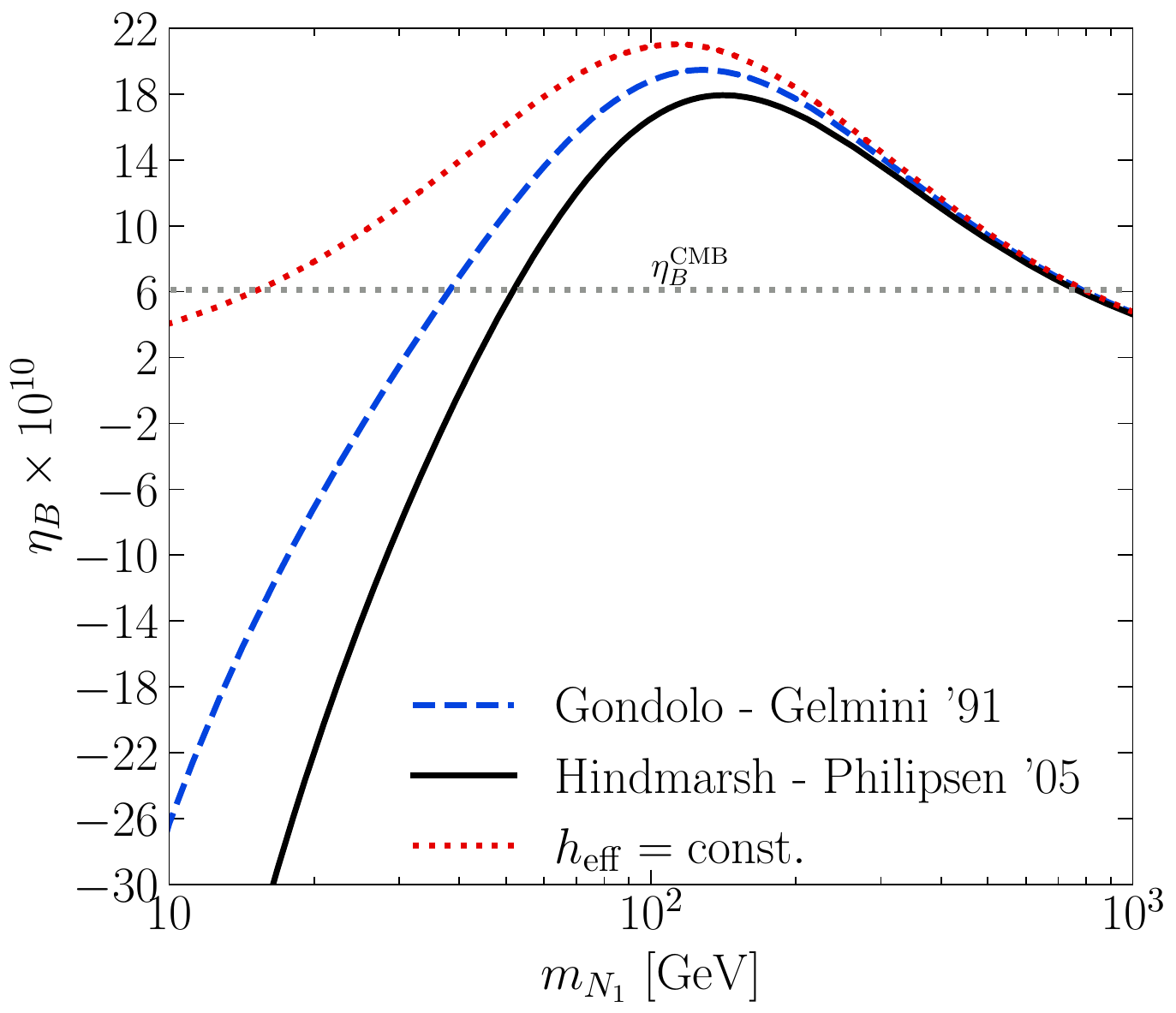}
		\caption{}
		\label{fig:heff_effect}
	\end{subfigure}
	\begin{subfigure}[]{0.48\textwidth}
		\includegraphics[width=1.\textwidth]{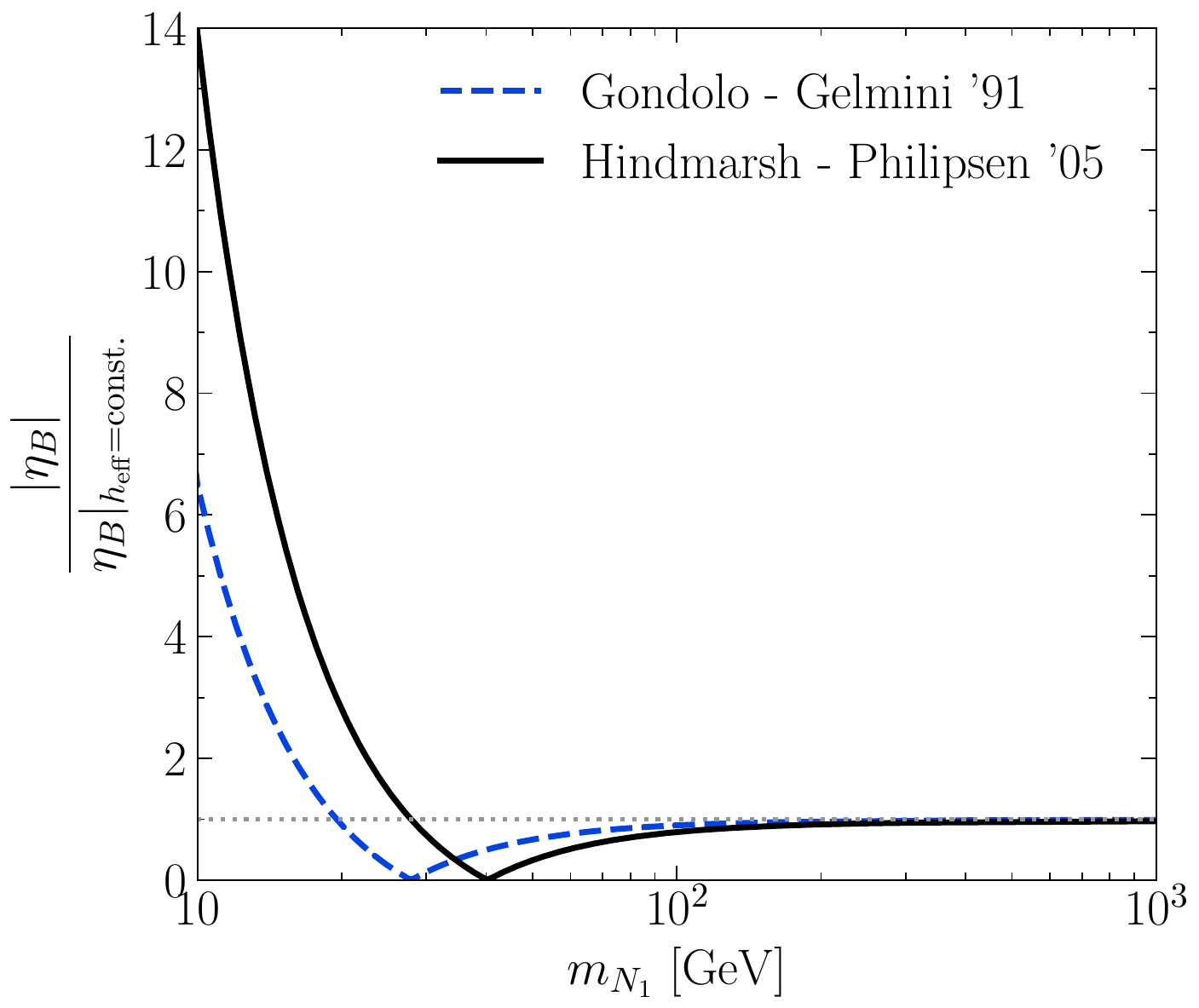}
		\caption{}
		\label{fig:heff_ratios}
	\end{subfigure}
	\centering
    \caption{\textit{Left panel:} The generated $\eta_B$ for
      $|\h^{\nu}_{ij}| \approx 3 \times 10^{-4}$ in the tri-resonant
      scenario as a function of $m_{N_1}$ for $\heff$ as given
      in~\cite{Hindmarsh:2005ix} (black),~\cite{Gondolo:1990dk}
      (blue), and taking $\heff = {\rm const.} \approx 105$ (red). The
      grey dotted line shows the observed baryon asymmetry, $\eta_B^{\rm CMB} = 6.104 \times
      10^{-10}$. \textit{Right panel:} The ratio of $|\eta_B|$ with
      varying $\heff$. The black (blue) line corresponds
      to~\cite{Hindmarsh:2005ix} (\hspace{1sp}\cite{Gondolo:1990dk}),
      with respect to $\heff = {\rm const.}$} 
    \label{fig:heff_var}
\end{figure}

We highlight the impact of the variations in the relativistic degrees of freedom in Figure~\ref{fig:heff_var}. From these figures, it is clear that when we utilise models with sub-TeV masses, the variations in the relativistic degrees of freedom may result in drastically different values for the observed BAU. In particular, we call attention to the fact that for singlet neutrino masses below $100 \;\GeV$, the observed BAU may take on negative values once the variations in the degrees of freedom are accounted for.

As a concluding remark, we note that this effect is stable under perturbations of the initial conditions since the solutions of the Boltzmann equations quickly reach attractor solutions. Moreover, we expect that this behaviour would pervade even with the inclusion of additional CP-violating phenomena, such as coherent heavy neutrino oscillations.

\section{Results}\label{sec:Res}
\begin{figure}[t!]
	\begin{subfigure}[]{0.48\textwidth}
		\includegraphics[width=1.\textwidth]{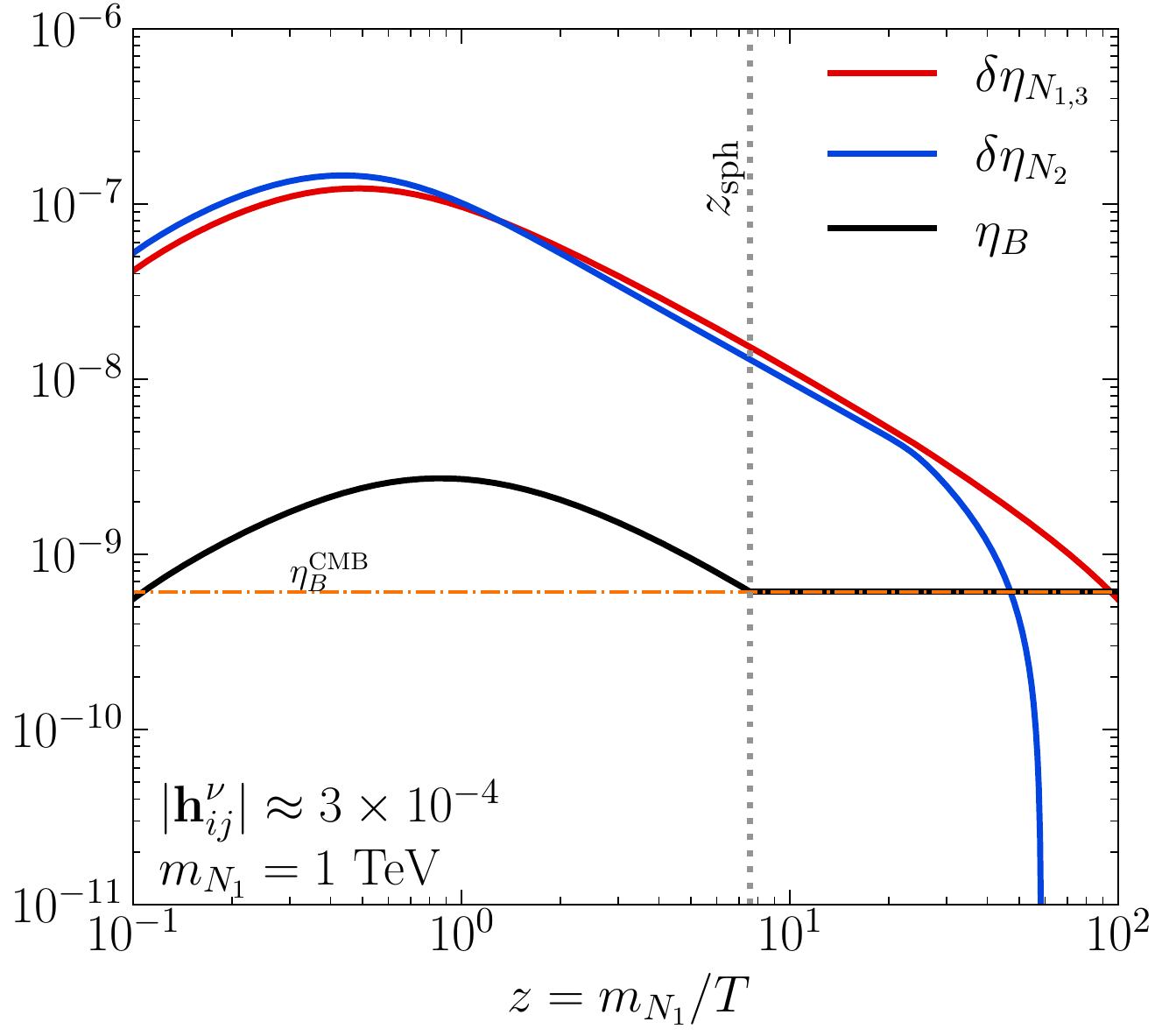}
		\caption{}
		\label{fig:evolution_high}
	\end{subfigure}
	\begin{subfigure}[]{0.48\textwidth}
		\includegraphics[width=1.\textwidth]{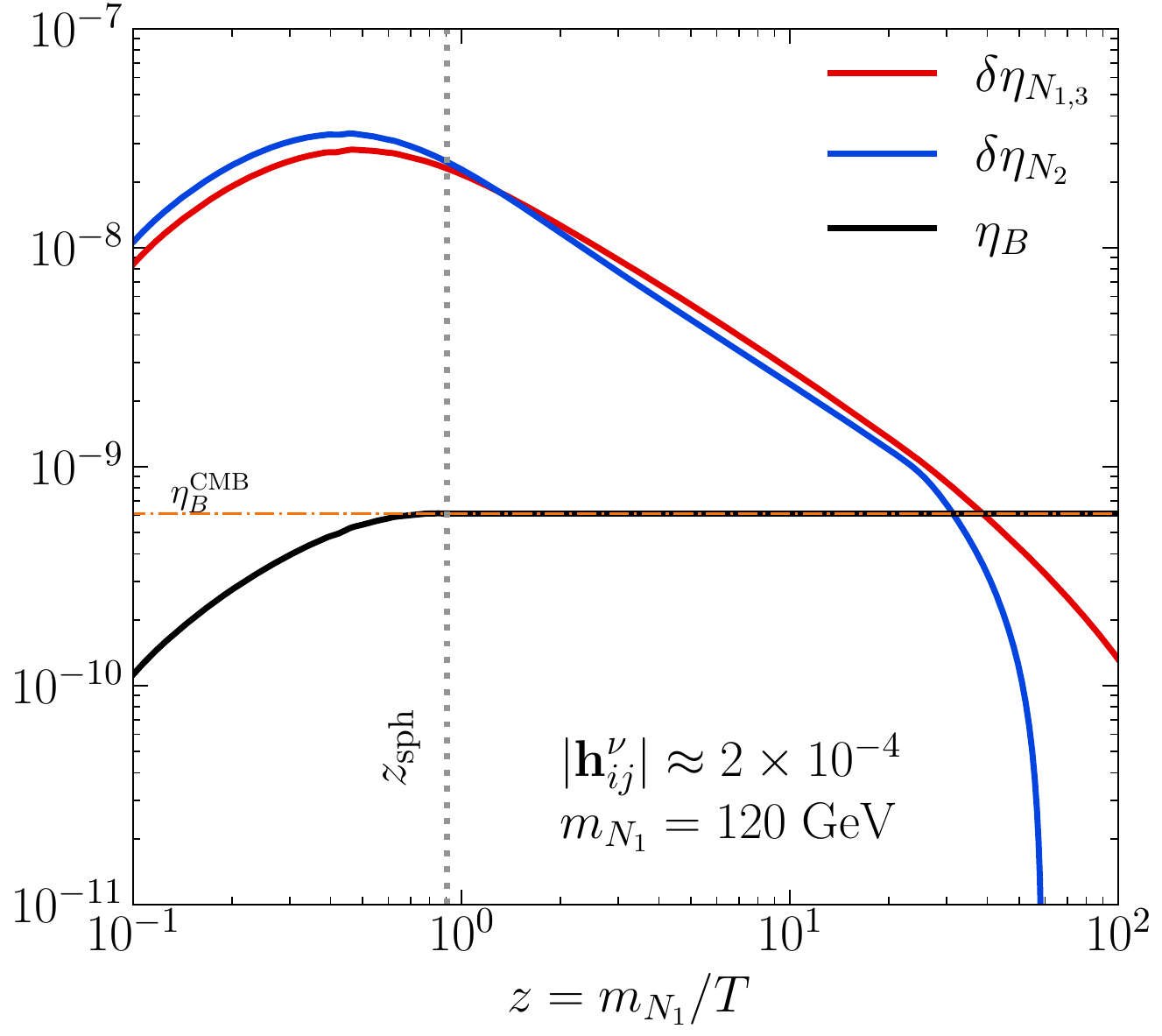}
		\caption{}
		\label{fig:evolution_low}
	\end{subfigure}
	\centering
    \caption{Evolution plots for the deviation from equilibrium of the neutrino
      densities $\delta\eta_{N_\alpha}$ (blue and red solid
      lines) and the baryon asymmetry $\eta_B$ (black
      solid line). Input parameters for the mass of the lightest
      singlet neutrino, and the scale of the Yukawa coupling can be
      seen on each panel, with the grey (dotted) line indicating the
      value $z=\zsph$ at which the sphaleron processes freeze out. The
      orange dot-dashed line indicates the observed value of the
      baryon asymmetry of $\eta^{\rm CMB}_B = 6.104\times 10^{-10}$.} 
    \label{fig:evol_plots}
\end{figure}

We now present the numerical solutions of the Boltzmann equations given in (\ref{eq:BEN}) and (\ref{eq:BEL}). We assume a tri-resonant mass spectrum for the singlet neutrinos and a democratic $a = b = c$ structure for the Yukawa couplings. We consider masses in the range $40 \; \GeV$ to $1 \; \TeV$, as numerical solutions below $40 \; \GeV$ are more limited due to the neglect of thermal masses, which may cause phase space suppression. In addition, a study at these low masses may require the inclusion of additional CP-violating effects, such as coherent oscillations of singlet neutrinos.

The inclusion of scattering processes generates a delay in the maximum of the baryon asymmetry evolution. As a result, the evolution of the baryon asymmetry becomes dependent on the mass scale of the singlet neutrinos. In Figure~\ref{fig:evol_plots}, we analyse this phenomenon. In both panels, we take the initial conditions $\delta \eta_{N_\alpha}(z_0) = 0$ and $\eta_L(z_0) = 0$, with $z_0 = 10^{-2}$, although due to the heavily attractive nature of the solution, the results remain unchanged for any other sensible choice of the initial conditions. Figure~\ref{fig:evolution_high} considers neutrinos of mass scale $m_{N_1} = 1 \; \TeV$ and Yukawa couplings $|\mathbf{h}_{ij}^\nu| = 3\times 10^{-4}$. It can be seen in this figure that for $\TeV$ scale neutrinos, the baryon asymmetry reaches a maximum value before a significant amount is washed out prior to the sphaleron freeze-out at $T_\textrm{sph}$. Conversely, Figure~\ref{fig:evolution_low} shows the evolution for neutrinos of mass $m_{N_1} = 120 \; \GeV$ and Yukawa couplings of size $|\mathbf{h}_{ij}^\nu| = 2\times 10^{-4}$. In this figure, we see that the generation of the BAU occurs at the maximum of the evolution. In general, light singlet neutrino masses results in the generated BAU freezing out earlier in the evolution. Finally, we observe that in both of the panels in Figure~\ref{fig:evol_plots}, at high values of $z$, there is a significant difference between the evolution of $\delta \eta_{N_2}$ and $\delta \eta_{N_{1,3}}$. This is due to the significantly higher CP asymmetry associated with $N_2$, as highlighted in Figure~\ref{fig:cp_asymmetry}.

\begin{figure}[t!]
\centering
\begin{subfigure}[]{0.49\textwidth}
	\includegraphics[width=1.0\textwidth]{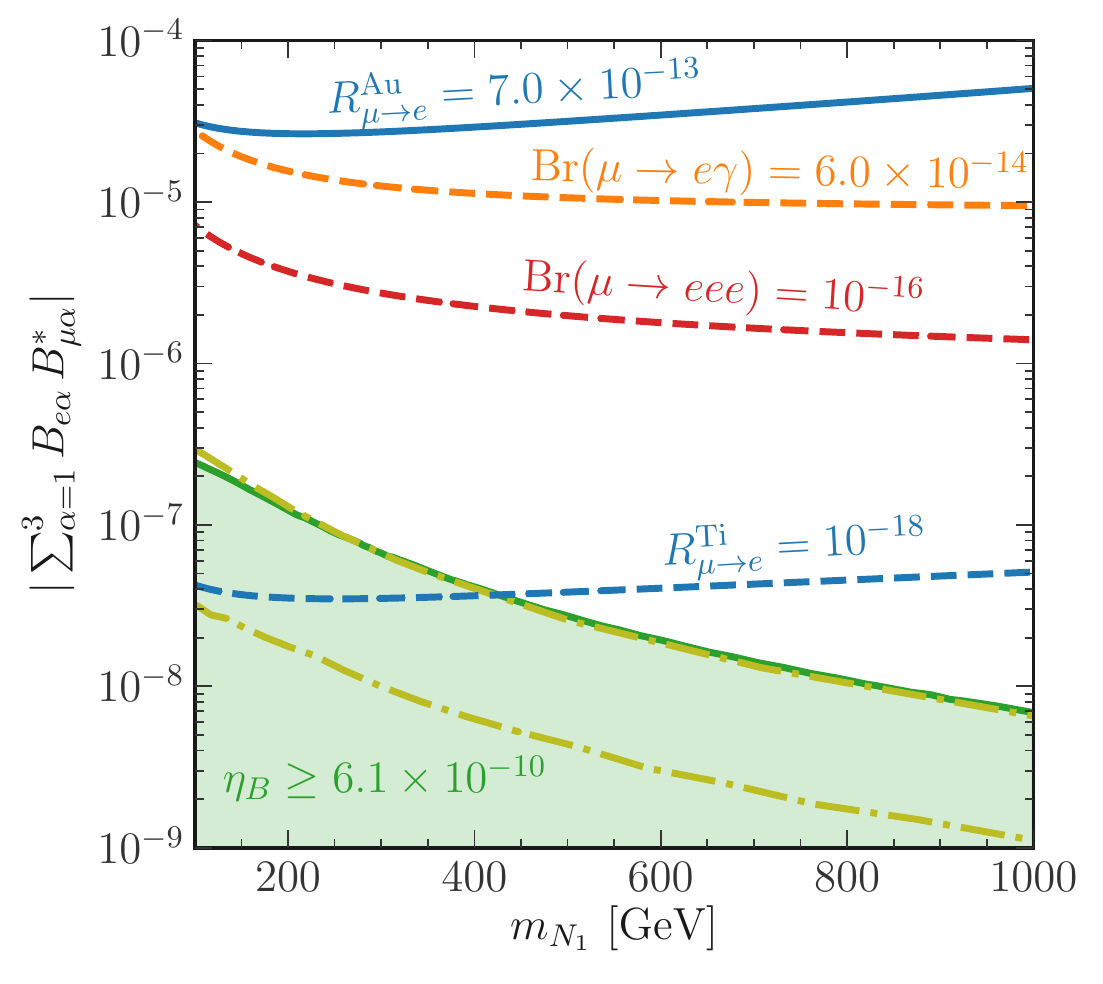}
	\caption{}
	\label{fig:lfv_lims}
\end{subfigure}
\begin{subfigure}[]{0.49\textwidth}
	\includegraphics[width=1.0\textwidth]{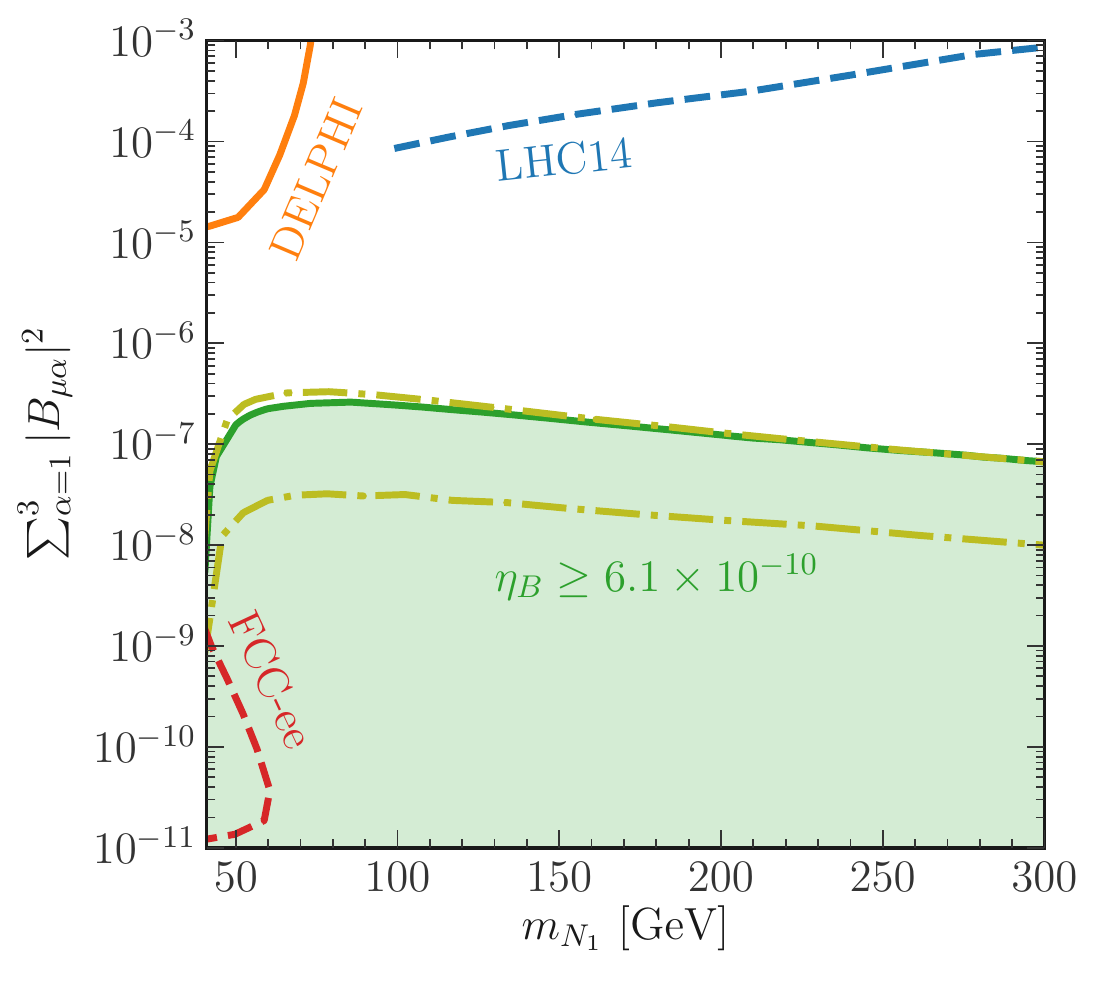}
	\caption{}
	\label{fig:collider_lims}
\end{subfigure}
\caption{Parameter space for the TRL model, including current limits
  (solid lines) and projected sensitivities of future experiments
  (dashed lines). \textit{Left panel:} Projected sensitivities of cLFV
  searches for $\mu\rightarrow e\gamma$ (orange dashed line),
  $\mu\rightarrow eee$ (dashed red line), coherent $\mu\rightarrow e$
  conversion in titanium (dashed blue line), and in gold (solid
  blue line). \textit{Right panel:} Projected
  sensitivities for collider searches at LHC$14$ (blue dashed line),
  FCC-ee (red dashed line), and current limits from DELPHI (orange
  solid line). The green region in these panels
  indicates points in the parameter space where successful leptogenesis is possible, and the green solid line corresponds to the points that reproduce exactly the
  observed value for a tri-resonant model. 
  The upper and lower yellow dot-dashed lines were obtained by scaling the total CP asymmetry $\delta_T$ by a factor of 2 and 0.1, respectively, then matching the observed baryon asymmetry. These lines represent an uncertainty estimate in the calculation of the solid green line due to the oscillations of singlet neutrinos.}
\label{fig:parameter_space}
\end{figure}

Figure~\ref{fig:parameter_space} shows the parameter space on the $\sum_\alpha B_{l\alpha} B^*_{k\alpha}$ vs. $m_{N_1}$ plane. As before, we assume a democratic flavour structure with $a=b=c$ and a tri-resonant mass spectrum. Moreover, we take the initial conditions $\delta \eta_{N_\alpha}(z_0) = 0$ and $\eta_L(z_0) = 0$, with $z_0 = 10^{-2}$. We highlight the region in which successful generation of the BAU is possible, with the solid green line indicating points in the parameter space where the generated BAU is equal to the observed value $\eta_B^\textrm{CMB} = 6.104 \times 10^{-10}$. Points within the green-shaded region may be made to match $\eta_B^\textrm{CMB}$ by softly relaxing the tri-resonant condition, and hence this region also permits the successful generation of the BAU.

In Figure~\ref{fig:parameter_space}, the yellow dashed lines represent bounds when additional sources of CP asymmetry are included. The upper dashed line is obtained by scaling up the total CP asymmetry by a factor of 2, and the lower dashed line is obtained by scaling down the CP asymmetry by a factor of 10. These represent the theoretical uncertainty due to the neglect of coherent oscillation effects between heavy neutrinos. These estimates were generated by assuming that the CP asymmetry from coherent oscillations is additive to the CP asymmetry arising through singlet neutrino mixing. However, there may be constructive or destructive interference between these two effects, highlighting the current lack of consensus regarding whether mixing and oscillations are distinct phenomena or whether mixing is contained within the oscillation formalism. Consequently, the numerical results from a detailed study containing both effects may not be as extreme as the bounds presented in Figure~\ref{fig:parameter_space}.

Figure~\ref{fig:lfv_lims} compares the available parameter space with sensitivity limits on current cLFV experiments involving muons. In particular, we consider coherent muon to electron transitions within nuclei, as well as $\mu \to e\gamma$ and $\mu \to eee$ experiments. As may be seen in this figure, the only experiment which may probe the parameter space of successful leptogenesis is the coherent $\mu\to e$ transition in Titanium at PRISM~\cite{BARLOW201144}

Figure~\ref{fig:collider_lims} considers the projected and current limits for various collider experiments. The estimate denoted by LHC$14$
(blue dashed line) presents conservative projections LHC
with $300~{\rm fb}^{-1}$ data operating at $\sqrt{s} =
14~\TeV$ for the sensitivity to the process $pp\rightarrow N\ell^\pm jj$~\cite{Deppisch:2015qwa,Dev:2013wba}. The orange solid line represents 95\% C.L. limits found by comparing LEP data with
the prediction for signals of decaying heavy neutrinos that are
produced via $Z\rightarrow N\nu_L$~\cite{DELPHI:1996qcc} at DELPHI. Similar
limits have been derived by the L3
collaboration~\cite{L3:1992xaz}. The red dashed line shows the Future Circular Collider (FCC)
sensitivity to the same signals
for electron-positron collisions assuming the normal order of the light
neutrino spectrum, and considering the lifetime of the heavy
neutrinos~\cite{Blondel:2014bra}.

\section{Conclusions}
We have showcased a class of leptogenesis models characterised by their $\mathbb{Z}_6$ or $\mathbb{Z}_3$ symmetries. These models offer naturally light SM neutrino masses with large CP-violating phases. When utilised in a tri-resonant framework, this model can fully saturate the available CP asymmetry. We have highlighted how the full and consistent incorporation of a third singlet neutrino can lead to significantly higher scales of CP asymmetry when compared with the bi-resonant approximation commonly utilised in the literature.

Furthermore, we have presented a complete set of Boltzmann equations, accounting for various effects such as varying degrees of freedom and chemical potential corrections. In addition, we have included scattering terms up to $\Delta L =2$ processes with proper RIS subtraction. In our analysis, we have explicitly demonstrated the importance of proper implementation of the variation of the degrees of freedom since this feature can have a significant impact on the generated BAU.

In addition, we have illustrated that an enhanced parameter space is possible when a tri-resonant mass spectrum is considered when compared with the expectations from a typical seesaw model. While the parameter space found is out of range for many current experiments, it is still possible for certain mass ranges to be probed, particularly through $\mu \to e$ conversion in Titanium at PRISM or through collision experiments at the FCC. Due to the democratic structure of the Tri-RL models, flavour effects will not be significant, and hence the results we provide give an upper bound on the scale of the neutrino Yukawa couplings. However, in principle, we may expand this parameter space through the inclusion of additional phenomena, such as coherent oscillations and supersymmetry (SUSY).

\section{Acknowledgements}
\noindent The work of AP and DK is supported in part by the Lancaster-Manchester-Sheffield Consortium for Fundamental Physics under STFC Research Grant ST/T001038/1. The work of PCdS is funded by Agencia Nacional de Investigaci\'{o}n y Desarrollo (ANID) through the Becas Chile Scholarship No. 72190359. TM acknowledges support from the STFC Doctoral Training Partnership under STFC training grant 
 ST/V506898/1.

\bibliography{bibs-refs}{}

\providecommand{\href}[2]{#2}\begingroup\raggedright\begin{thebibliography}{10}

\bibitem{Planck:2018vyg}
{\bf Planck} Collaboration, N.~Aghanim et~al., {\it {Planck 2018 results. VI.
  Cosmological parameters}},  {\em Astron. Astrophys.} {\bf 641} (2020) A6,
  [\href{http://arxiv.org/abs/1807.06209}{{\tt arXiv:1807.06209}}]. [Erratum:
  Astron.Astrophys. 652, C4 (2021)].

\bibitem{Fields:2019pfx}
B.~D. Fields, K.~A. Olive, T.-H. Yeh, and C.~Young, {\it {Big-Bang
  Nucleosynthesis after Planck}},  {\em JCAP} {\bf 03} (2020) 010,
  [\href{http://arxiv.org/abs/1912.01132}{{\tt arXiv:1912.01132}}]. [Erratum:
  JCAP 11, E02 (2020)].

\bibitem{Ahmad:2001an}
{\bf SNO} Collaboration, Q.~R. Ahmad et~al., {\it {Measurement of the rate of
  $\nu_e+d \to p+p+e^-$ interactions produced by $^8B$ solar neutrinos at the
  Sudbury Neutrino Observatory}},  {\em Phys. Rev. Lett.} {\bf 87} (2001)
  071301, [\href{http://arxiv.org/abs/nucl-ex/0106015}{{\tt nucl-ex/0106015}}].

\bibitem{Ahmad:2002jz}
{\bf SNO} Collaboration, Q.~R. Ahmad et~al., {\it {Direct evidence for neutrino
  flavor transformation from neutral current interactions in the Sudbury
  Neutrino Observatory}},  {\em Phys. Rev. Lett.} {\bf 89} (2002) 011301,
  [\href{http://arxiv.org/abs/nucl-ex/0204008}{{\tt nucl-ex/0204008}}].

\bibitem{Fukuda:1998mi}
{\bf Super-Kamiokande} Collaboration, Y.~Fukuda et~al., {\it {Evidence for
  oscillation of atmospheric neutrinos}},  {\em Phys. Rev. Lett.} {\bf 81}
  (1998) 1562--1567, [\href{http://arxiv.org/abs/hep-ex/9807003}{{\tt
  hep-ex/9807003}}].

\bibitem{Minkowski:1977sc}
P.~Minkowski, {\it {$\mu \to e\gamma$ at a Rate of One Out of $10^{9}$ Muon
  Decays?}},  {\em Phys. Lett.} {\bf 67B} (1977) 421--428.

\bibitem{GellMann:1980vs}
M.~Gell-Mann, P.~Ramond, and R.~Slansky, {\it {Complex Spinors and Unified
  Theories}},  {\em Conf. Proc.} {\bf C790927} (1979) 315--321,
  [\href{http://arxiv.org/abs/1306.4669}{{\tt arXiv:1306.4669}}].

\bibitem{Yanagida:1979as}
T.~Yanagida, {\it {Horizontal gauge symmetry and masses of neutrinos}},  {\em
  Conf. Proc.} {\bf C7902131} (1979) 95--99.

\bibitem{Mohapatra:1979ia}
R.~N. Mohapatra and G.~Senjanovic, {\it {Neutrino Mass and Spontaneous Parity
  Nonconservation}},  {\em Phys. Rev. Lett.} {\bf 44} (1980) 912.

\bibitem{Sakharov:1967dj}
A.~D. Sakharov, {\it {Violation of CP Invariance, C asymmetry, and baryon
  asymmetry of the universe}},  {\em Pisma Zh. Eksp. Teor. Fiz.} {\bf 5} (1967)
  32--35.

\bibitem{FukYan:1986}
M.~Fukugita and T.~Yanagida, {\it {Baryogenesis without grand unification}},
  {\em Phys. Lett.} {\bf B174} (1986) 45.

\bibitem{Buchmuller:2004nz}
W.~Buchmuller, P.~Di~Bari, and M.~Plumacher, {\it {Leptogenesis for
  pedestrians}},  {\em Annals Phys.} {\bf 315} (2005) 305--351,
  [\href{http://arxiv.org/abs/hep-ph/0401240}{{\tt hep-ph/0401240}}].

\bibitem{daSilva:2022mrx}
P.~C. da~Silva, D.~Karamitros, T.~McKelvey, and A.~Pilaftsis, {\it
  {Tri-resonant leptogenesis in a seesaw extension of the Standard Model}},
  {\em JHEP} {\bf 11} (2022) 065, [\href{http://arxiv.org/abs/2206.08352}{{\tt
  arXiv:2206.08352}}].

\bibitem{DOnofrio:2014rug}
M.~D'Onofrio, K.~Rummukainen, and A.~Tranberg, {\it {Sphaleron Rate in the
  Minimal Standard Model}},  {\em Phys. Rev. Lett.} {\bf 113} (2014), no.~14
  141602, [\href{http://arxiv.org/abs/1404.3565}{{\tt arXiv:1404.3565}}].

\bibitem{MEG:2016leq}
{\bf MEG} Collaboration, A.~M. Baldini et~al., {\it {Search for the lepton
  flavour violating decay $\mu ^+ \rightarrow \mathrm {e}^+ \gamma $ with the
  full dataset of the MEG experiment}},  {\em Eur. Phys. J. C} {\bf 76} (2016),
  no.~8 434, [\href{http://arxiv.org/abs/1605.05081}{{\tt arXiv:1605.05081}}].

\bibitem{MEGII:2018kmf}
{\bf MEG II} Collaboration, A.~M. Baldini et~al., {\it {The design of the MEG
  II experiment}},  {\em Eur. Phys. J. C} {\bf 78} (2018), no.~5 380,
  [\href{http://arxiv.org/abs/1801.04688}{{\tt arXiv:1801.04688}}].

\bibitem{BARLOW201144}
R.~Barlow, {\it {The PRISM/PRIME Project}},  {\em Nuclear Physics B -
  Proceedings Supplements} {\bf 218} (2011), no.~1 44--49. Proceedings of the
  Eleventh International Workshop on Tau Lepton Physics.

\bibitem{Pilaftsis:1991ug}
A.~Pilaftsis, {\it {Radiatively induced neutrino masses and large Higgs
  neutrino couplings in the standard model with Majorana fields}},  {\em Z.
  Phys. C} {\bf 55} (1992) 275--282,
  [\href{http://arxiv.org/abs/hep-ph/9901206}{{\tt hep-ph/9901206}}].

\bibitem{Pilaftsis:1998pd}
A.~Pilaftsis, {\it {Heavy Majorana neutrinos and baryogenesis}},  {\em Int. J.
  Mod. Phys. A} {\bf 14} (1999) 1811--1858,
  [\href{http://arxiv.org/abs/hep-ph/9812256}{{\tt hep-ph/9812256}}].

\bibitem{Pilaftsis:1997jf}
A.~Pilaftsis, {\it {CP violation and baryogenesis due to heavy Majorana
  neutrinos}},  {\em Phys. Rev. D} {\bf 56} (1997) 5431--5451,
  [\href{http://arxiv.org/abs/hep-ph/9707235}{{\tt hep-ph/9707235}}].

\bibitem{Pilaftsis:2003gt}
A.~Pilaftsis and T.~E.~J. Underwood, {\it {Resonant leptogenesis}},  {\em Nucl.
  Phys. B} {\bf 692} (2004) 303--345,
  [\href{http://arxiv.org/abs/hep-ph/0309342}{{\tt hep-ph/0309342}}].

\bibitem{Pilaftsis:2005rv}
A.~Pilaftsis and T.~E.~J. Underwood, {\it {Electroweak-scale resonant
  leptogenesis}},  {\em Phys. Rev. D} {\bf 72} (2005) 113001,
  [\href{http://arxiv.org/abs/hep-ph/0506107}{{\tt hep-ph/0506107}}].

\bibitem{Deppisch:2010fr}
F.~F. Deppisch and A.~Pilaftsis, {\it {Lepton Flavour Violation and
  $\theta_{13}$ in Minimal Resonant Leptogenesis}},  {\em Phys. Rev. D} {\bf
  83} (2011) 076007, [\href{http://arxiv.org/abs/1012.1834}{{\tt
  arXiv:1012.1834}}].

\bibitem{Pilaftsis:1997dr}
A.~Pilaftsis, {\it {Resonant CP violation induced by particle mixing in
  transition amplitudes}},  {\em Nucl. Phys. B} {\bf 504} (1997) 61--107,
  [\href{http://arxiv.org/abs/hep-ph/9702393}{{\tt hep-ph/9702393}}].

\bibitem{Branco:1986gr}
G.~C. Branco, L.~Lavoura, and M.~N. Rebelo, {\it {Majorana Neutrinos and {CP}
  Violation in the Leptonic Sector}},  {\em Phys. Lett. B} {\bf 180} (1986)
  264--268.

\bibitem{Yu:2020gre}
B.~Yu and S.~Zhou, {\it {Sufficient and Necessary Conditions for CP
  Conservation in the Case of Degenerate Majorana Neutrino Masses}},  {\em
  Phys. Rev. D} {\bf 103} (2021), no.~3 035017,
  [\href{http://arxiv.org/abs/2009.12347}{{\tt arXiv:2009.12347}}].

\bibitem{KUZMIN198536}
V.~Kuzmin, V.~Rubakov, and M.~Shaposhnikov, {\it {On anomalous electroweak
  baryon-number non-conservation in the early universe}},  {\em Physics Letters
  B} {\bf 155} (1985), no.~1 36--42.

\bibitem{Hindmarsh:2005ix}
M.~Hindmarsh and O.~Philipsen, {\it {WIMP dark matter and the QCD equation of
  state}},  {\em Phys. Rev. D} {\bf 71} (2005) 087302,
  [\href{http://arxiv.org/abs/hep-ph/0501232}{{\tt hep-ph/0501232}}].

\bibitem{PhysRevD.42.3344}
J.~A. Harvey and M.~S. Turner, {\it {Cosmological baryon and lepton number in
  the presence of electroweak fermion-number violation}},  {\em Phys. Rev. D}
  {\bf 42} (Nov, 1990) 3344--3349.

\bibitem{Gondolo:1990dk}
P.~Gondolo and G.~Gelmini, {\it {Cosmic abundances of stable particles:
  Improved analysis}},  {\em Nucl. Phys. B} {\bf 360} (1991) 145--179.

\bibitem{Deppisch:2015qwa}
F.~F. Deppisch, P.~S. Bhupal~Dev, and A.~Pilaftsis, {\it {Neutrinos and
  Collider Physics}},  {\em New J. Phys.} {\bf 17} (2015), no.~7 075019,
  [\href{http://arxiv.org/abs/1502.06541}{{\tt arXiv:1502.06541}}].

\bibitem{Dev:2013wba}
P.~S.~B. Dev, A.~Pilaftsis, and U.~K. Yang, {\it {New Production Mechanism for
  Heavy Neutrinos at the LHC}},  {\em Phys. Rev. Lett.} {\bf 112} (2014), no.~8
  081801, [\href{http://arxiv.org/abs/1308.2209}{{\tt arXiv:1308.2209}}].

\bibitem{DELPHI:1996qcc}
{\bf DELPHI} Collaboration, P.~Abreu et~al., {\it {Search for neutral heavy
  leptons produced in Z decays}},  {\em Z. Phys. C} {\bf 74} (1997) 57--71.
  [Erratum: Z.Phys.C 75, 580 (1997)].

\bibitem{L3:1992xaz}
{\bf L3} Collaboration, O.~Adriani et~al., {\it {Search for isosinglet neutral
  heavy leptons in Z0 decays}},  {\em Phys. Lett. B} {\bf 295} (1992) 371--382.

\bibitem{Blondel:2014bra}
{\bf FCC-ee study Team} Collaboration, A.~Blondel, E.~Graverini, N.~Serra, and
  M.~Shaposhnikov, {\it {Search for Heavy Right Handed Neutrinos at the
  FCC-ee}},  {\em Nucl. Part. Phys. Proc.} {\bf 273-275} (2016) 1883--1890,
  [\href{http://arxiv.org/abs/1411.5230}{{\tt arXiv:1411.5230}}].

\end{thebibliography}\endgroup


\providecommand{\href}[2]{#2}\begingroup\raggedright\endgroup
\bibliographystyle{JHEP}
\end{document}